\documentstyle[epsf]{aa}
\begin{document}
\title{Accurate radius and mass of the transiting exoplanet OGLE-TR-132b \thanks{Based on observations
           collected with the FORS2 imager at the VLT/UT4 Yepun telescope
           (Paranal Observatory, ESO, Chile) in the DDT programme 273.C-5017A.}}
\author{Moutou C.$^1$, Pont F.$^2$, Bouchy F.$^1$, Mayor M.$^2$}
\offprints{Claire.Moutou@oamp.fr}
\institute{$^1$ LAM, Traverse du Siphon, BP8, Les Trois Lucs, 13376 Marseille
  cedex 12, France\\
$^2$ Observatoire de Gen\`eve, 51 Chemin des Maillettes,1290 Sauverny, Switerland}
\date{Received date / accepted date}
   \authorrunning{C. Moutou et al.}
   \titlerunning{Accurate radius and mass of OGLE-TR-132b}
\abstract{ 
The exoplanet OGLE-TR-132b belongs 
to the new class of very hot giant planets, together with OGLE-TR-56b
and OGLE-TR-113b, detected by
their transits. Recently, radial velocity measurements provided a planetary
mass estimate for OGLE-TR-132b. The planet parameters, however, were poorly constrained,
because of the very shallow transit in the OGLE light curve.
In this letter, based on 
new VLT/FORS2 photometric follow-up of OGLE-TR-132 of unprecedented quality (1.2
millimagnitude relative photometry), we confirm
the planetary nature of the orbiting object, and we derive an accurate
measurement of its radius and mass: 1.13 $\pm$ 0.08 $R_J$ and 1.19 
$\pm$ 0.13  $M_J$. The refined ephemeris of OGLE-TR-132 transits is $T_0 = $ 2453142.5888
and $P =$ 1.689857 days.
      }
\maketitle
\section{Introduction}
The rich complementarity of transit and radial-velocity methods in the
detection and study of extrasolar planets has opened the way to characterizing
some of these planets.
It is now possible in particular to investigate 
the mass-radius relationship of extrasolar giant planets.
The first transiting exoplanet HD209458b (Charbonneau et al. 2000,
Henry et al. 2000) has been extensively observed, up to the recent detection of its
exosphere (Vidal-Madjar et al. 2003). Several transiting 
hot Jupiter candidates were detected by OGLE (Optical Gravitational Lensing
Experiment, Udalski et al. 2002a, 2002b, 2003) and three of them were confirmed by
radial-velocity follow-up: OGLE-TR-56b (Konacki et al. 2003),
OGLE-TR-113b (Bouchy et al. 2004, Konacki et al. 2004) and
OGLE-TR-132b (Bouchy et al. 2004, hereafter referenced as "BPS"). The low orbital periods of these
three planets (less than 1.7 days) make them extreme cases of the "hot Jupiter" class.
They have radii and masses consistent with one another 
and markedly higher density than HD209458b (Brown et al. 2001). 

The transit dip of OGLE-TR-132 in the OGLE light curve is very shallow 
compared to the mean photometric error bars ($d\sim 0.008, \sigma \sim 0.006 $), providing 
little constraints on the parameters of the transit, in particular the 
impact parameter. BPS
show that the data was consistent with a central transit of a $R\sim 1.3
R_\odot$ star by a $r \sim 1 R_J$ planet, but also with a high-latitude transit of a much 
larger star by a larger planet. 

The reality of the transit and radial velocity signals themselves were not entirely beyond doubt.
OGLE-TR-132 belongs to a "supplement" of very shallow signals detected in the OGLE data with
the BLS algorithm (Kov\'acs et al., 2002) which
also includes at least one object with no radial velocity signal (OGLE-TR-131, BPS) that
can be suspected to be a false positive. Moreover, BPS estimate that the radial velocity
signal of OGLE-TR-132 has a 3\% chance of being due to random fluctuation in the velocity residuals.

In this letter, we present a much more accurate light curve of the transit 
of OGLE-TR-132, obtained with the FORS2 imager on the VLT. The new data unambiguously
confirm the reality of both the photometric transit and radial velocity signal, and 
allow precise contraints to be put on the shape and phasing of the transit, 
therefore yielding much improved radius and mass determinations for the planetary companion.
\section{Observations and data reduction}
The observations were obtained during 4 hours on May 16th, 2004 on the FORS2
camera of VLT/UT4 (programme 273.C-5017A). In total, 281 short exposures
(15-20 sec) in the
$R_{special}$ filter were acquired, in a 3.4'x3.4' field of view around OGLE-TR-132. 
Two images in
the Bessel $V$ filter were acquired before and after the long $R_{special}$ sequence,
for colour calibration. We used FORS2 with the high-resolution collimator and
a 2x2 pixel binning in order to get both spatial sampling and short readout
time. The pixel size is 0.12".
The atmospheric conditions during the observed sequence were excellent and
stable: the average seeing at zenith is 0.55" with up to 20\% fluctuations, and
transparency was high and stable.
The airmass of the field grows from 1.26 to 1.58 during the sequence.
We obtained approximately one photometric measurement per minute with exposure times
of 15 or 20 seconds;
the sequence was scheduled following recent radial velocity measurements which
suggested a possible shift of up to 4 hours from the OGLE ephemeris (BPS). 

\begin{figure}
\centerline{\epsfxsize=8cm\epsfbox{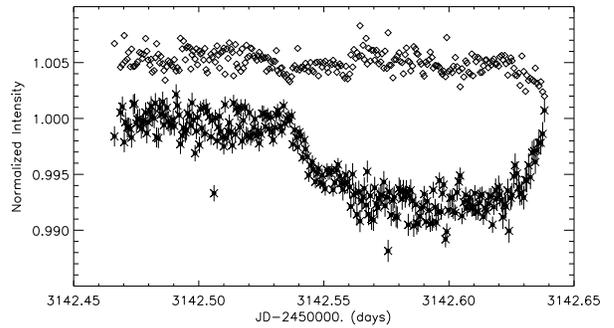}}
\caption{Our VLT/FORS light curve for the observed transit of
  OGLE-TR-132. The top curve (losanges) is the mean light curve obtained from 20 stars in
  the field (shifted by 0.005 for clarity), from which the target light curve (bottom)
  is corrected.}
\label{transit}
\end{figure}

The frames were debiassed and flatfielded with the standard ESO pipeline. 
In the following, only the chip 1 where OGLE-TR-132 lies is mentionned (half of the field of view).
Differential photometry was performed with the image subtraction algorithm
(Alard \& Lupton, 1998 and Alard, 2000). The combination of the three best-quality images 
led to the creation of the reference frame.
After an astrometric interpolation of all images (3rd-degree polynomial
fitting of the distortions), all images are then scaled 
to the reference using adapted kernels to compensate 
for the varying seeing and background conditions. 
Aperture photometry is then performed with the DAOPHOT package in 
IRAF (Stetson 1987) to both the
reference frame and subtracted images. 
Different apertures were
tested for the extraction, and the best compromise between including all the flux of the 
object and minimizing the sky background was found for an aperture of 10 pixel
radius.
The resulting light curve of
OGLE-TR-132 is shown on Figure~\ref{transit} (data are available electronically). 
Several tens of stars 
of similar colour and magnitude than OGLE-TR-132 were measured for
comparison and twenty of them were selected to produce an average light curve 
(under conditions of low intrinsic
variability, clean sky area, large brightness without
saturation and low blending). This comparison light curve is then subtracted
to the light curve of OGLE-TR-132 to remove residual systematics. 
A small second-order residual was found in the sky background substraction between the 15-second and
the 20-second exposures. This residual, amounting to 0.4 mmag for the 
optimal aperture of 10 pixel radius, was removed in the 
final light curve.

OGLE-TR-132 has a close visual companion about 1" to the east. We also 
checked that this companion was not variable: the
variations of the 20-times-fainter companion are of the order of 2\%, due to
residuals in the PSF wing of OGLE-TR-132. Were the companion responsible for the
observed transit, it should be variable by $\simeq$15\%, which is not observed.

\begin{figure}
\centerline{\epsfxsize=8cm\epsfbox{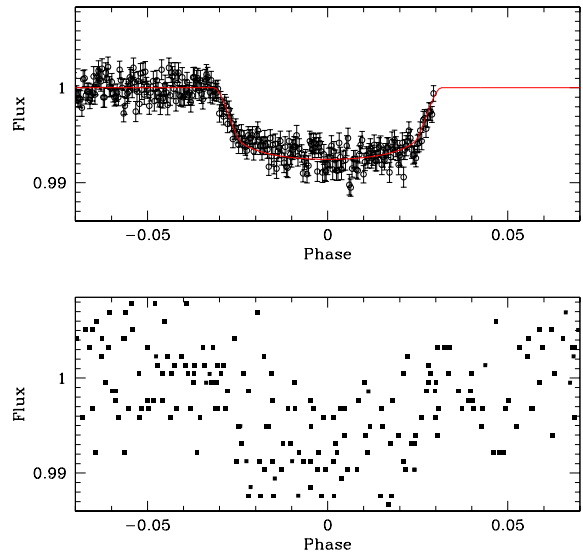}}
\caption{{\bf Top:} The best-fit transit curve is shown together with the
  phased FORS2 data (outliers were removed by 3$\sigma$ clipping). 
  The error bars show the photon noise. {\bf Bottom:} 
  On the same scale, the OGLE data from 11 individual transits (data from
  Udalski et al. 2003).}
\label{trfit}
\end{figure}
\section{Results}
\subsection{Light curve analysis}
The dispersion in the resulting light curve for our comparison stars is 
very close to the photon noise limit. The dispersion of the light curve of OGLE-TR-132 before the transit
is 1.19 mmag, and the mean photon noise is 1.04 mmag. The dispersion of the residuals
during the transit after the transit fit (see below) is 1.14 mmag. We can therefore
estimate by quadratically substracting the photon noise that the systematic effects
are smaller than 0.7 mmag. We also checked that colour effects are
negligible. 
Incidentally, this excellent accuracy shows what can be achieved under good 
conditions at the VLT on a $I=15.72$ magnitude star.

\subsection{Transit fitting}
A transit light curve was fitted by least-squares to the photometric
data with five free
parameters: the radius ratio $r/R$, the transit duration $d$, the 
impact parameters $b$, the period $P$ and the epoch $T_0$. A quadratic limb darkening\footnote{The data
is actually good enough for a linear limb darkening coefficient to be left as a free parameter.
Doing this yields $u=0.6 \pm 0.1$.} 
with $u_1=u_2=0.3$ was used (applicable for a late F dwarf according to Barban et
al. 2003). The  expected transit shape was computed according to Mandel \&
Agol (2002).
The fit was performed in two stages: the transit period was first constrained with the 
combined OGLE and FORS2 data, which spans over two years. 
The period
was then fixed and the other parameters were fitted on the FORS2 data only.
A 3-$\sigma$ clipping was applied to the data to remove outliers (8 points out of 281) and the
uncertainties were scaled from the photon noise with a factor 1.19 for the final fit so that 
the reduced $\chi^2$ be equal to the number of degrees of freedom.
The uncertainties on the parameters were computed from the reduced $\chi^2$ of
the combined fit.

The results are given in Table~\ref{table1} and the fit is shown on Figure~\ref{trfit}. The data
place very tight constraints on the transit shape. The duration of the transit (defined as the crossing
time of the center of the planet through the star's disc) is defined within a few seconds. The shapes of 
the ingress and egress are also clearly defined, showing the transit to occur at medium latitude on the star.
Note that due to the geometry of the problem, the uncertainty distributions are not always symmetrical. For 
instance, the one-sigma interval for the impact parameter $b$ is 0.44-0.64, but the two-sigma interval
is 0.0 - 0.71, so that a central transit is excluded at the two-sigma level by
the data\footnote{The uncertainty
distribution on the physical parameters (mass and radius) is, however, 
reasonably symmetrical (e.g. a central transit would yield $r=0.91 R_J$)}.
\begin{table}
\begin{tabular}{l l }
\hline
Period [days]& 1.689857 $\pm$ 0.000006\\
Transit epoch [JD]& 2453142.58884 $\pm$ 0.00009 \\
Radius ratio & 0.0812 $\pm$ 0.0017\\
Impact parameter & 0.57 $\pm$ 0.10\\
Transit duration [day] & $0.09476 ^{+0.00027}_{-0.00007}$ \\
 & \\
Temperature of primary [K] $^a$ & 6411 $\pm$ 179 \\
 V-R & 0.58 $\pm$ 0.05\\
$M_I$ [mag] & 3.05 $\pm$ 0.21 \\
age [Gyr]& 0 - 1.4 \\
Distance [pc] & 2500 $\pm$ 250\\
Primary mass [$M_\odot$]&  1.35 $\pm$ 0.06\\
Primary radius [$R_\odot$]& 1.43 $\pm$ 0.10\\
 & \\
Planet mass [$M_J$]& 1.19 $\pm$ 0.13\\
Planet radius [$R_J$]& 1.13 $\pm$ 0.08\\
Planet density [$g.cm^{-3}$]& 1.02 $\pm$ 0.33 \\ \hline
\end{tabular}
\caption{Parameters for the observed transit, the star OGLE-TR-132 and its planetary companion.
$^a$: from BPS. }
\label{table1}
\end{table}
\subsection{Radius and mass determination}
\begin{figure}
\centerline{\epsfxsize=8cm\epsfbox{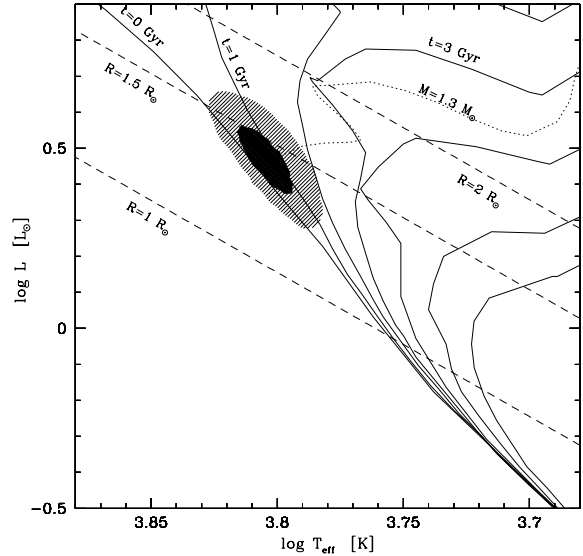}}
\caption{Position of OGLE-TR-132 in the temperature-luminosity diagram. The 68\% (cross-hatched) and 90\% (hatched) 
contours are indicated for the  least-squares described in the text. Some landmarks from the Girardi et al. (2002) stellar
evolution models are indicated: isochrones of the $z=0.05$ ([Fe/H]$\sim +0.4$) models for ages 0,1,2,3,5,10,15 Gyr;
 evolution track for $z=0.03$ and $M=1.3 M_\odot$. Lines of constant radius at 1, 1.5 and 2 $R_\odot$ are also indicated.}
\label{figcmd}
\end{figure}

The radius and mass of OGLE-TR-132 and of its planetary companion can be computed
by combining the constraints from the transit curve and from the spectroscopy. Given the very short
period, a circular Keplerian orbit can safely be assumed.
There are two constraints on the stellar radius and mass:\\
- The transit duration in phase, which is proportional to $R\, M^{-1/3}$ for a circular Keplerian orbit.\\
- The spectroscopic determination of temperature, metallicity 
and gravity by BPS.\\
These two constraints were combined by generalised least-squares, assuming that
the parameters are confined in the ($T_{eff}$, $\log g$, [Fe/H], $M$, $R$) space in the
sub-manifold defined by an interpolation between the Girardi et al. (2002) stellar
evolution models\footnote{The interpolation was done with the IAC-star software of A. Aparicio, priv. comm.
}. The results are given in Table~\ref{table1} and illustrated in Fig.~\ref{figcmd}.

This procedure, in addition to $M$ and $R$, also yields estimates of the stellar age, absolute
magnitude and intrinsic colours: $\tau = 0 - 1.4 $ Gyr, $M_I=3.05\pm 0.21$ mag, $(V-R)_0=0.28\pm 0.02$ mag.
Using these values and our measurement of $(V-R)$, we derive $a_v=1.43\pm0.22$ and a distance of 2500$\pm$250 pc for OGLE-TR-132.

\begin{figure}
\centerline{\epsfxsize=8cm\epsfbox{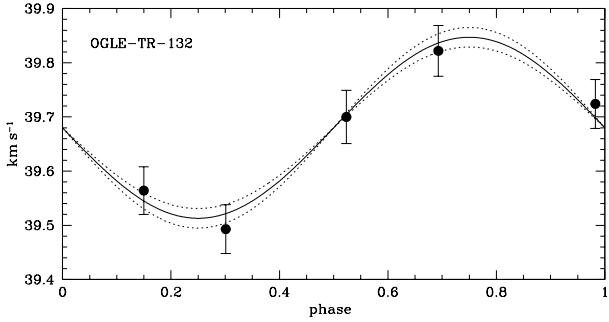}}
\caption{The radial-velocity measurements of OGLE-TR-132 obtained with
  VLT/FLAMES (BPS), with the re-evaluated fit. The dotted lines
  correspond to fit curves for lower and upper 1-$\sigma$ intervals in
  semi-amplitude $K$.}
\label{Vr}
\end{figure}

Our improved period and epoch determination compared to Udalski et al. (2003) 
also leads to an improved determination of the semi-amplitude of the radial velocity orbit, $K$. 
We fitted a sinusoidal orbit 
by least-squares to the radial velocity data with the new period and epoch. 
The result, $K= 0.167 \pm 0.018$,  is plotted in Fig.~\ref{Vr}. The new period 
leads to a significant improvement over the inital value of BPS 
and to a larger value of $K$. The phasing of the transit is now in much closer
agreement with the variation of the radial velocity. We use this new value in the 
determination of $m$ from $M$. The mass $m$ of the 
planetary companion can be computed from  the mass $M$ of 
the star and  $K$ (for a circular orbit, $K\sim m\, (m+M)^{2/3}$).

The radius $r$ of the planetary companion can be computed from 
the radius $R$ of  the star and the radius ratio $r/R$. 
Our values for $r$ and $m$ are given in Table~\ref{table1}.

\section{Discussion and conclusion}
The planetary transit is thus detected without ambiguity in our FORS2 data set and well
sampled in time. From a transit fitting and a new analysis of the VLT/FLAMES
radial-velocity curve, we could determine
the planet radius and mass with a high accuracy (Table~\ref{table1}).
The remaining limitation on the planetary radius precision now results 
in comparable levels from the systematic residuals 
in the photometry, the uncertainties in the parent star spectroscopic 
parameters, and the accuracy of stellar evolution models.

\begin{figure}
\centerline{\epsfxsize=8cm\epsfbox{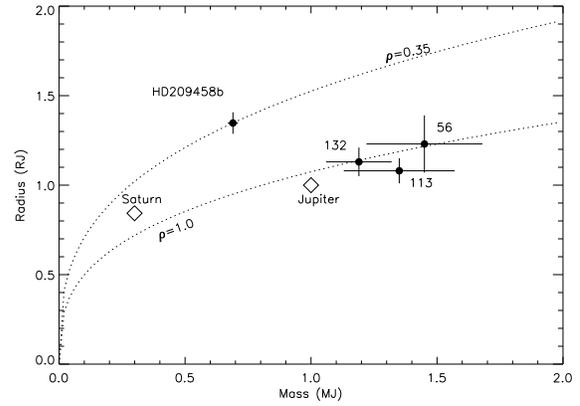}}
\caption{The mass-radius relationship of the four known transiting
planets (from  Brown et al., 2001, Cody \& Sasselov 2002, BPS,
Torres et al. 2004 and this work). OGLE-TR-132 is coded "132", etc...
Iso-density lines are also drawn (in units gcm$^{-3}$). Jupiter and Saturn are
shown for comparison.}
\label{data}
\end{figure}

The determination of planet radius and mass give a good estimate of the mean density of OGLE-TR-132b, showing that
it is in line with other close-in hot Jupiters, OGLE-TR-56b and OGLE-TR-113b. We confirm the significance of the
difference with HD209458b, which has a markedly lower density.  
Figure~\ref{data} shows the mass-radius relationship of the four
extrasolar planets where both parameters are measured.

Following Lammer et al. (2003) and Baraffe et al. (2004), we may
 estimate the mass loss of the OGLE-TR-132
planet due to extreme UV stellar radiation. It would be of the
order of 10$^{-10}$ $M_J/yr$ if the parent star was of solar type,
but it is more likely greater, since the star is hotter
and younger (and thus more luminous in the XUV domain)
than the Sun. This minimum mass loss estimate is yet
twice as large as the mass
loss of OGLE-TR-56b and four times larger than for HD209458b.
We could expect strong signatures of an evaporated exosphere from
OGLE-TR-132, larger than those observed in the spectrum of
HD209458b (Vidal-Madjar et al., 2003, 2004).\\
Further transit searches and their radial-velocity follow-up will
probably provide additional candidates of great interest.


\end{document}